\def\vec#1{{\rm\bf #1}}

\documentclass[11pt]{article} 
\usepackage{graphics}
\usepackage[dvips]{graphicx}
\usepackage{amssymb}
\usepackage{amsmath}
\usepackage{vmargin}
\begin{document}

\title{The stress transmission universality classes of periodic granular arrays}
\author{R. C. Ball \\ Department of Physics, University of Warwick, \\ Coventry  CV4 7AL, UK\\
and \\
D. V. Grinev \\ Cavendish Laboratory, University of Cambridge, \\ Madingley Road, Cambridge CB3 OHE} 
\maketitle

\begin{abstract}

The transmission of stress is analysed for static periodic arrays of
rigid grains,  with perfect and zero friction.  For minimal coordination
number (which is sensitive to friction,  sphericity and dimensionality),
  the stress distribution is soluble without reference to the
corresponding displacement fields. In non-degenerate cases, the
constitutive equations are found to be simple linear in the stress
components. The corresponding coefficients depend crucially upon
geometrical disorder of the grain contacts. 
\end{abstract}

\section{Introduction}

Granular media present a fascinating array of physical problems \cite{Jaeger} and  are key to a number of important technologies \cite{Ennis}.
Distribution of stress and statistics of force fluctuations in static granular arrays show puzzling properties \cite{stress}. 
For instance, photoelastic visualization experiments \cite{photo} show that stresses in granular media concentrate along "stress paths" or "force chains". 
These are filamentary configurations of grains which  carry disproportionally large amount of the total force and give rise to the phenomenon of 
jamming \cite{jamming}. The intergranular contact forces determine the bulk properties (e.g. the load bearing capability) of granular materials  
\cite{Guyon}. Recent experiments \cite{Mueth} and computer simulations \cite{Radjai} show that the distribution of forces where forces above 
the mean decay exponentiallly is a robust property of static granular media. Despite recent theoretical attempts \cite{Wittmer,Coppersmith,Edwards} 
and a vast engineering literature \cite {Nedderman,Sokolovskii} the transmission of stress and statistical properties of contact force distribution in 
granular media are still poorly understood at a fundamental level. 
This paper is concerned with a static granular material idealised as an
assembly of rigid grains,  not in general spherical and with either
zero or perfect (in the sense of infinite) friction.  Whilst
practical applications in soil mechanics and powder technology evidently 
depend on much greater levels of detail considered in the engineering
literature,  the rigid grain paradigm provides a crucial starting point
from which to appreciate the theoretical physics of the problem.
We show below that for minimal coordination number there exists an
analysis of stress which is independent of the analysis of strain which
would be implied by slight relaxation of the rigid grain assumption.  In
consequence our results may also have bearing on many computer simulations of
quasi-static interacting particles, such as applied to colloids \cite{colloid} or soils \cite{Wood}, 
and suggest that the detail of conservative repulsive force laws used may not always be significant.

\section{The Model}

Consider a static assembly of rigid grains of average coordination
number $z$,  in dimension $d$. The microscopic version of stress
analysis consists of
determining all of the intergranular forces, given the geometrical arrangement of grains and their contacts and the applied force and
torque loadings on each grain.  The number of unknowns per grain is $zd'/2$,  where $d'=d$ for grains with friction and $d'=1$ for grains
without,  and the required force and torque give $d + (d^2-d)/2$
constraints.  The minimal coordination number,  at which the equations
of force balance are directly soluble,  is thus given by $z_m = \frac{d^2+d}{d'}\,$.
Without friction this gives $z_m = 6,\, 12$ in $d= 2,\, 3$
respectively.  With friction it gives rather low values $z_m=3,\, 4$ respectively.
If we admit that grains have some compliance then the
intergranular forces calculated from Newton's equationsmust also be
consistent with being calculated from the displacements of grains. This implies $z d'$ constraints on the
particle displacements, which at minimal coordination is precisely
consistent with determining the center displacement and rotation of each
particle.  This situation is analogous to linear elasticity in two
dimensions, where stress can be analysed independent of the elastic
moduli for an isotropic material; subsequently the
displacement field can be recovered (using the moduli) and of course
this may be relevant for boundary conditions \cite{Landau}.  For the granular
material,  no assumption of linearity or isotropy is involved.
Equivalent arguments, without consideration of torques and rotations, go
through for frictionless spherical grains.  They exhibit minimal
coordination number equal to twice the dimension of space. 
In this paper we will elaborate in detail the constraint equations
satisfied by the stress tensor supported by simple periodic granular
arrays.  Here we argue how the key result, of simple linear equations of
constraint,  can be anticipated on quite general grounds. We consider a
periodic array with $M$ grains per unit cell and correspondingly
$M z_m/2$ intergranular contacts. 
The number of strictly periodic ($\vec{k}=\vec{0}$)  solutions for the
intergranular forces dictates how many linearly independent stress
components the powder can support macroscopically.  The general stress
field of our periodic array will be decomposable over Bloch wave
solutions $\propto e^{i \vec{k}.\vec{r}}$, and within these it is the
periodic solutions with wavevector $\vec{k}=\vec{0}$ which
correspond to macroscopically
uniform stress. 
For a periodic solution the intergranular forces will be constrained by
$(M-1)d$ equations of force balance rather than $M d$, because no
intergranular force can apply a net force to the whole assembly.  Unless
there
is accidental degeneracy (see end of this section) there will be no such
mitigation of
the number of torque constraints at $M(d + (d^2-d)/2)$.
  The general result is then that at minimal coordination we have
precisely $d$ degrees of freedom corresponding to macroscopically uniform stress, to
be determined macroscopically by the $d$ macroscopic (continuum)
equations of force balance $\vec{\nabla}.\sigma=\vec{0}$, where $\sigma$
is the stress tensor.  The number of equations restricting the form of
macroscopic stress supported, equivalent to a constitutive equation, is
$(d^2-d)/2$, equivalent to one in $d=2$ and three in $d=3$.  These
equations will be developed explicitly for special cases below. 
The anomalous case is,  unfortunately,  the one case previously
considered \cite{Ball}:  minimal periodic lattices of spherical grains with friction.
These are the honeycomb and diamond lattices in $d=2,\, 3$ dimensions
respectively,  which have two grains per unit cell related to each other
by a reflection symmetry.  The reflection symmetry means that if the
torque on one particle is balanced,  then so it must also be on the
other.  As there is no corresponding reduction in the number of
independent inter-particle forces,  the number of degrees of freedom for
solutions corresponding to macroscopic stress is increased to
$(d^2+d)/2$,  corresponding to a full set of symmetric stress tensors.
As shown in Ref.\cite{Ball}, the constitutive equation then  takes
differential form and has to be found from leading $\vec{k}$-dependent
behaviour.
In three dimensions (and higher) the existence of intermediate classes
of behaviour can be conjectured,  where some but not all of the
constitutive equations are differential in form.  Candidate geometries
include simple arrays of ellipsoids,  but will not be presented here.

\section{Frictionless aspherical grains in $d=2$}

\begin{figure}[h] 
\begin{center} 
\resizebox{5cm}{!}{\rotatebox{270}{\includegraphics{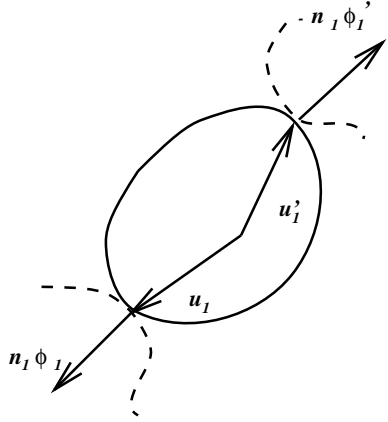}}}
\caption{Normal forces $\vec{n}\phi$ and $-\vec{n}\phi'$ at the points of contact}
\label{slipgrain} 
\end{center} 
\end{figure}

Consider a periodic triangular array of smooth grains with the geometry
of the contacts and their normals as shown in Figure \ref{slipgrain}.
Contact 1 is at positition $\vec{u}_1$ relative to the grain center, with surface outward normal $\vec{n}_1$ subject to a force $\phi_1 \vec{n}_1$.
 Note that periodicity requires what would be contact 4 to have normal
$-\vec{n}_1$, to which end we have labelled it $1'$, and similarly for the
other opposed pairs of contacts. 
Balance of force around the grain,  subject to external force $\vec{F}$,
requires:

\begin{equation}\label{frank1}
\vec{F} + (\phi_1 - \phi'_1)\vec{n}_1 + (\phi_2 -
\phi'_2)\vec{n}_2
 + (\phi_3 - \phi'_3)\vec{n}_3 = \vec{0}
\end{equation}
and the (tensorial) force moment around the grain is given by:

\begin{equation}\label{frank2} 
 S= (\phi_1 \vec{u}_1 - \phi'_1 \vec{u}'_1)\vec{n}_1 +
 (\phi_2 \vec{u}_2 - \phi'_2 \vec{u}'_2)\vec{n}_2 + 
 (\phi_3 \vec{u}_3 - \phi'_3 \vec{u}'_3)\vec{n}_3
\end{equation}
In terms of $S$,  the balance of torque on the grain requires that $S$
be symmetric, and
the macroscopic stress tensor is given by:

\begin{equation}\label{frank3} 
\sigma = \frac{1}{V}\sum_{g \in V}\, S_{g}
\end{equation}
For Bloch wave solutions of wavevector $\vec{k}$ we have $\phi'_1 =
\phi_1 e^{i\vec{k}.(\vec{u_{1}}'-\vec{u_{1}})}$ and similarly for $\phi'_2$ and
$\phi'_3$.  Then to leading order in $k$ we obtain:

\begin{equation}\label{frank4} 
 S= \phi_1 \vec{a}_1 \vec{n}_1 +
 \phi_2 \vec{a}_2 \vec{n}_2 + 
 \phi_3 \vec{a}_3 \vec{n}_3 + \hbox{ order }k
\end{equation}
where $\vec{a}_1 = \vec{u}_1 - \vec{u}'_1$ {\it etc.} are the lattice
vectors,  and:

\begin{equation}\label{frank5}
 \vec{F} + i\vec{k}.S + \hbox{ order }k^2  = \vec{0}
\end{equation}
This last equation is just macroscopic force balance, equivalent to
$\vec{\nabla}.\sigma + \vec{f} = \vec{0}$,  where $\vec{f}$ is the
external force per unit volume.
The expression for $S$ together
with the constraint that it be symmetric restricts the form of the
stress tensor.  It is readily verified that the lattice vectors satisfy
a triangle equality $\vec{a}_1 + \vec{a}_2 + \vec{a}_3 =\vec{0}$,  and
we can without loss of generality rescale the lengths of the normals to
obtain a further triangle of vectors
 $\vec{m}_1 + \vec{m}_2 + \vec{m}_3 =\vec{0}$,  where $\vec{m}_1 \propto
\vec{n}_1$ {\it etc}.  Then it is not difficult to check that
$\vec{a}_1\times S \times \vec{m}_2 =\vec{a}_2\times S \times
\vec{m}_1$; the two further identities that this appears to offer by
changing numerical labels are easily shown to be equivalent to it.  Thus
the
constitutive equation restricting the stress tensor can be written
explicitly as:

\begin{equation}\label{frank6}
\hbox{Trace }(P.S) = 0
\end{equation}
governed by an order parameter $P$ given by:

\begin{equation}\label{frank7}
 P = R.\vec{m}_2 R.\vec{a}_1 - R.\vec{m}_1 R.\vec{a}_2
\end{equation}
where $R$ is a rotation through $\pi/2$. 
It is interesting that the order parameter $P$ is intrinsically chiral,
with its sign sensitive to how we index contacts round the grain.  If
the isotropic part of $P$ happens to vanish,  then our constitutive
equation reduces to Fixed Principal Axes (FPA) ansatz \cite{Wittmer}.  However as all the vectors
$\vec{a}_1$, $\vec{a}_2$, $\vec{m}_1$, $\vec{m}_2$ are in principle
independent of each other,  there is no reason to expect:

\begin{equation}\label{frank8} 
\hbox{Trace}P
= \vec{m}_2.\vec{a}_1 - \vec{m}_1.\vec{a}_2 = 0 
\end{equation}
and FPA ansatz remains a
special case.

\section{Frictionless aspherical grains in $d=3$}

The simplest regular array to consider here is (affinely distorted)
face-center cubic, duly giving the minimal coordination number of 12.
Opposing pairs of contacts are mutually constrained as in the two
dimensional case, and as wavevector $k$ approaches zero we find the
force moment tensor given by a sum over the six pairs as:

\begin{equation}\label{frank9} 
 S= \sum_{i=1}^6{\phi_i \vec{a}_i \vec{n}_i}
\end{equation}
Making $S$ symmetric determines three of the $\phi_i$,  so that we are
left with just three degrees of freedom for the symmetric stress, as
expected.
The connectivity of the lattice ensures that the six lattice vectors
span the edges of a tetrahedron,  and there is one normal vector
corresponding to each edge.  Although it is not in general possible to
scale the normals to fit the edges of a tetrahedron,  we can scale them
in usefully analogous way. The three lattice vectors meeting at any 
vertex have, given appropriate ordering conventions, the property that
$\vec{a}.\vec{a}'\times\vec{a}" = 6V$,  where the constant $V$ is the
volume of the tetrahedron.  The six normal vectors can be rescaled as
$\vec{m}_i \propto \vec{n}_i$ so
that any corresponding set of three obeys the analogous result
$\vec{m}.\vec{m}'\times\vec{m}" = 1$.
Given the above conventions it is not difficult to see that if edges
1,2,3,4 form a circuit enclosing triangles 1,2,5 and 3,4,5 then
$(\vec{a}_1\times\vec{a}_2).S.(\vec{m}_3\times\vec{m}_4) =
(\vec{a}_3\times\vec{a}_4).S.(\vec{m}_1\times\vec{m}_2)$
as both sides are equal to $6V\phi_6$ with appropriate scale
factor. In this way we can obtain six constraint
equations of the form:

\begin{equation}\label{frank10} 
(\vec{a}_1\times\vec{a}_2 \, \vec{m}_3\times\vec{m}_4 -
\vec{a}_3\times\vec{a}_4 \, \vec{m}_1\times\vec{m}_2):S = 0
\end{equation}
of which only three are linearly independent of each other.
For frictionless spheres the simplest periodic lattice of minimal
coordination number, $2d$, is a cubic array,  sheared over in the
general case. For this arrangement it is
obvious that the system can support only normal stresses along the
symmetry axes.  When these are at right angles this
corresponds to Fixed Principle Axis behaviour,  but not otherwise.

\section{Grains with friction}

For grains with friction,  the simplest periodic arrays require at
least two grains per unit cell because of the relativley low minimal
coordination number.  The honeycomb and diamond lattices (for two and
three dimensions respectively) have already been discused in \cite{Ball},
but with an assumption of symmetry between the two grains of the unit
cell: this is that the intracell grain contact lies at
the centroid of the intercell grain contacts. Here we show that if the
intracell contact is displaced by a vector $\vec{c}$,  a simple linear
rather than differential constitutive equation results.
Working for simpicity in two dimensions and focussing on $\vec{k}=\vec{0}$
solutions, the force moment (proportional to stress tensor) of the unit
cell can be written as:

\begin{equation}\label{frank11} 
 S=\vec{a}_1 \vec{f_1} + \vec{a}_2 \vec{f}_2
\end{equation} 
where $\vec{f_i}$ is the force across the intercell contacts spanned by lattice
vector $\vec{a}_i$. From this we can find the $\vec{f}_i$ in terms of $S$
and hence expressions for the torque applied to each grain in the unit
cell.  The sum of these torques vanishes when $S$ is symmetric, but the
difference between them is given (at $\vec{k}=\vec{0}$) by:

\begin{equation}\label{frank12}  
\Delta G = {\vec{c}\times S\times
(\vec{a}_1-\vec{a}_2) \over \vec{a}_1\times\vec{a}_2}
\end{equation} 
resulting in a constitutive equation

\begin{equation}\label{frank13}  
\vec{c}\times \sigma \times (\vec{a}_1-\vec{a}_2) = 0
\end{equation} 
In three dimensions the diamond lattice calculation is analogous, with
three intercell grain contacts spanning three lattice vectors,  and the
resulting constitutive equation is:

\begin{equation}\label{frank14}  
\vec{c}\times
\sigma.(\vec{a}_1\times\vec{a}_2+\vec{a}_2\times\vec{a}_3+
\vec{a}_3\times\vec{a}_1) = \vec{0}
\end{equation}
In both cases,  when the grains become equivalent and $\vec{c}=\vec{0}$
one must work to higher order in the wavevector $\vec{k}$,  leading to a
differential constitutive equation as shown in \cite{Ball}.

\section{Discussion}

In this paper we have shown how the geometrical arrangement of grains
and their contacts,  assumed known,  directly restricts the form of
stress tensor that the material can support.  In principle one can only
apply forces to a sample,  and the restrictions found on the stress
tensor are
no more than that required to solve for the resulting stress
distribution.
Independent of friction, dimensionality or even whether the particles are
spheres, we find a general universality class $GL$ of mechanical
constitutive equation which is simple linear in the stress components. The
coefficients of this are sensitive to the intergranular geometry to a much
finer level than simple objects such as the fabric tensor.  Two further
universality classes have been found.  The first is where by symmetry of
the granular array there is no coupling (of the constitutive equations) to
$\hbox{Trace}\,\sigma$, equivalent to FPA ansatz behaviour \cite{Wittmer}. 
The second class, DL, is where by internal symmetry of the granular array
some of the constitutive equations become differential\cite{Ball}.
All of the above behaviour is summed up by the general form:

\begin{equation}\label{frank15}  
 N:\sigma + \nabla.T:\sigma +  ... =0
\end{equation}

The nul space of $N$ acting on $\sigma$ corresponds to allowed macroscopic
stress
fields;  when this includes pure isotropic stress we have the special
class FPA.  When the nul space is degenerately large, $T$ becomes crucial
and we have DL behaviour:  the only instances of this which we explicitly
calculated had $N=0$,  but the existence of mixed cases in three
dimensions (and higher) is conjectured. In practice detail of intergranular contacts is not known in advance,
but should be deduced from the deposition history of the system. Truly
history dependent problems are outside the scope of this paper, but for
granular systems which have consolidated or sheared under the applied
loading, and for pseudo-elastic assemblies which have undergone
significant deformation and rearrangement under stress and/or flow, we
can consider the approximation that the current stress itself
influences the contact geometry. In two dimensions we then require to make one scalar equation out of the
stress tensor alone, and assuming it to be independent of the magnitude
of the stress this must reduce to a condition on the ratio of the
principal stress components, $\sigma_1/\sigma_2 = \hbox{constant }$. 
This is of precisely the same form as classical considerations of
limiting internal friction:

\begin{equation}\label{frank16}  
\mu =\frac{\sigma_{\hbox{shear}}}{\sigma_{\hbox{normal}}} = \frac{\sigma_{1}- \sigma_{2}}{2\sqrt{\sigma_{1}\sigma_{2}}}
\end{equation}

where $\mu$ is the coeffficient of shearing friction. In three dimensions we require three equations, which cannot be imposed
on the only two principle stress ratios available.  In this case,
therefore, it appears that we must have inescapably history-dependent
behaviour.  There might remain the possibility that the system selects
degenerate case configurations for which at least one of the
constitutive equations becomes differential in form,  but to achieve
this appears to place conditions on the sample history or,  leading to
contradiction,  the present stress tensor.
Work on disordered arrays of rigid grains is in progress \cite{disorder}.

\section{Acknowledgements}
The authors acknowledge valuable discussions with J.R.Melrose, C.Thornton,
R.Farr and T.A.Witten. D.V.G. acknowledges financial support from Shell
(Amsterdam), EPSRC (UK) and R.C.B. partial support from USNSF under grant PHY94-07194.

\end{document}